\documentclass[%
 preprint,
 amsmath,amssymb,
 aps,
]{revtex4-1}

\usepackage{graphicx}
\usepackage{dcolumn}
\usepackage{bm}

\begin{document}

\title{Propagating elastic vibrations dominate thermal conduction in amorphous silicon}

\author{Jaeyun Moon}
\affiliation{%
	Division of Engineering and Applied Science\\
	California Institute of Technology, Pasadena, California 91125,USA
}%
\author{Benoit Latour}
\affiliation{%
	Division of Engineering and Applied Science\\
	California Institute of Technology, Pasadena, California 91125,USA
}%

\author{Austin J. Minnich}%

\email{aminnich@caltech.edu}
\affiliation{%
	Division of Engineering and Applied Science\\
	California Institute of Technology, Pasadena, California 91125,USA
}%

\date{\today}

\begin{abstract}
	Thermal atomic vibrations in amorphous solids can be distinguished by whether they propagate as elastic waves or do not propagate due to lack of atomic periodicity. In a-Si, prior works concluded that non-propagating waves are the dominant contributors to heat transport, while propagating waves are restricted to frequencies less than a few THz and are scattered by anharmonicity. Here, we present a lattice and molecular dynamics analysis of vibrations in a-Si that supports a qualitatively different picture in which propagating elastic waves dominate the thermal conduction and are scattered by elastic fluctuations rather than anharmonicity. We explicitly demonstrate the propagating nature of vibration with frequency approaching 10 THz using a triggered wave computational experiment. Our work suggests that most heat is carried by propagating elastic waves in a-Si and demonstrates a route to achieve extreme thermal properties in amorphous materials by manipulating elastic fluctuations.

\end{abstract}

\pacs{Valid PACS appear here}
\maketitle


	Amorphous materials are of interest for a wide range of applications due to their low thermal conductivity \cite{foote_thermopile_2004,wingert_thermal_2016}. While in crystals heat is carried by propagating lattice waves, or phonons, in amorphous solids heat carriers are classified as propagons, diffusons, and locons depending on the degree of delocalization of the atomic vibration and its mean free path \cite{allen_diffusons_1999, feldman_thermal_1993}. 
	
	This classification has been widely used to analyze the vibrations responsible for thermal transport in amorphous materials, especially for pure a-Si. Numerical studies using equilibrium molecular dynamics (EMD) and lattice dynamics (LD) have attempted to determine the fraction of heat carried by each category of vibration \cite{he_heat_2011,larkin_thermal_2014}. While the general consensus is that diffusons carry the majority of the heat, prior works have reported that propagons may carry $20\sim50 \%$ of thermal conductivity in a-Si due to their long mean free paths \cite{allen_diffusons_1999,he_heat_2011}. Using normal-mode analysis, Larkin and McGaughey reported that propagons have a lifetime scaling of $\omega ^{-2}$ which suggests plane-wave-like propagation that is not affected by atomic disorder \cite{larkin_thermal_2014}. The normal mode lifetime analysis of Lv and Henry concluded that the phonon gas model is not applicable to amorphous materials \cite{lv_examining_2016}. Experimental works have qualitatively confirmed some of these predictions, particularly regarding the important contribution of propagons \cite{cahill_thermal_1994, sultan_heat_2013, zink_thermal_2006, braun_size_2016, kwon_unusually_2017,liu_high_2009}. For instance, Kwon et al. observed size effects in a-Si nanostructures, indicating the presence of propagons \cite{kwon_unusually_2017}.

	Despite these efforts, numerous puzzles remain. One discrepancy concerns the conclusion that the lifetimes of few THz vibrations are governed by anharmonicity \cite{larkin_thermal_2014}.  If that is the case, explaining the low thermal conductivity of a-Si is challenging because the same vibrations contribute 75 Wm\textsuperscript{-1}K\textsuperscript{-1} to thermal conductivity in c-Si. Accounting for the low thermal conductivity of a-Si only by changes in anharmonicity requires large increases in anharmonic force constants that would necessarily have effects on the heat capacity of a-Si that have not been observed \cite{zink_thermal_2006}. Along similar lines, if lifetimes of low frequency vibrations are governed by anharmonicity the reported thermal conductivities of films of the same thickness should be reasonably uniform, yet the data vary widely \cite{cahill_thermal_1994,zink_thermal_2006,liu_high_2009}. Overall, an unambiguous classification of the propagating nature and scattering mechanisms of vibrational modes transporting heat in amorphous solids is poorly developed, impeding efforts to synthesize, for example, novel materials with exceptionally low thermal conductivity.
	
	In this work, we address these questions using lattice and molecular dynamics to calculate dynamic structure factors and thermal transport properties of a-Si. Our analysis supports a qualitatively different picture of atomic vibrations in a-Si from the conventional one in which propagating elastic waves dominate the thermal conduction and are scattered by elastic fluctuations rather than anharmonicity. We explicitly demonstrate the propagating nature of waves with frequencies approaching 10 THz using a computational triggered wave analysis. Our work provides strong evidence that, unintuitively, elastic waves with frequencies up to 10 THz carry substantial heat in disordered media and also demonstrates a route to create materials with exceptional thermal properties by manipulating elastic fluctuations.     
	
	We used lattice and molecular dynamics to examine the atomic vibrations of various amorphous domains. The molecular dynamics calculations were performed using the Large-scale Atomic/Molecular Massively Parallel Simulator (LAMMPS) with a timestep of 0.5 fs \cite{plimpton_fast_1995}.  Periodic boundary conditions were imposed and the Stillinger-Weber interatomic potential was used \cite{stillinger_computer_1985}. The initial structure we considered contained 4096 atoms and was created by first melting crystalline silicon at 3500 K for 500 ps in an NVT ensemble. Next, the liquid silicon was quenched to 1000 K with the quench rate of 100 K/ps. The structures were annealed at 1000 K for 25 ns to reduce metastabilities \cite{moon_sub-amorphous_2016}. Finally, the domain was quenched at a rate of 100 K/ps to 300 K and equilibrated at 300 K for 10 ns in an NVT ensemble using a Nose-Hoover thermostat. The structure was then equilibrated at 300 K for 500 ps in an NVT ensemble. After an additional equilibration in an NVE ensemble for 500 ps, the heat fluxes were computed for 1.6 ns in the same NVE ensemble. We use Green-Kubo theory to compute the thermal conductivity of the structure to be 1.5 Wm\textsuperscript{-1}K\textsuperscript{-1}, a value that is consistent with prior works \cite{larkin_thermal_2014,lv_direct_2016,moon_sub-amorphous_2016}. 
	
\begin{figure}
	\centering
	\includegraphics[width=1\linewidth]{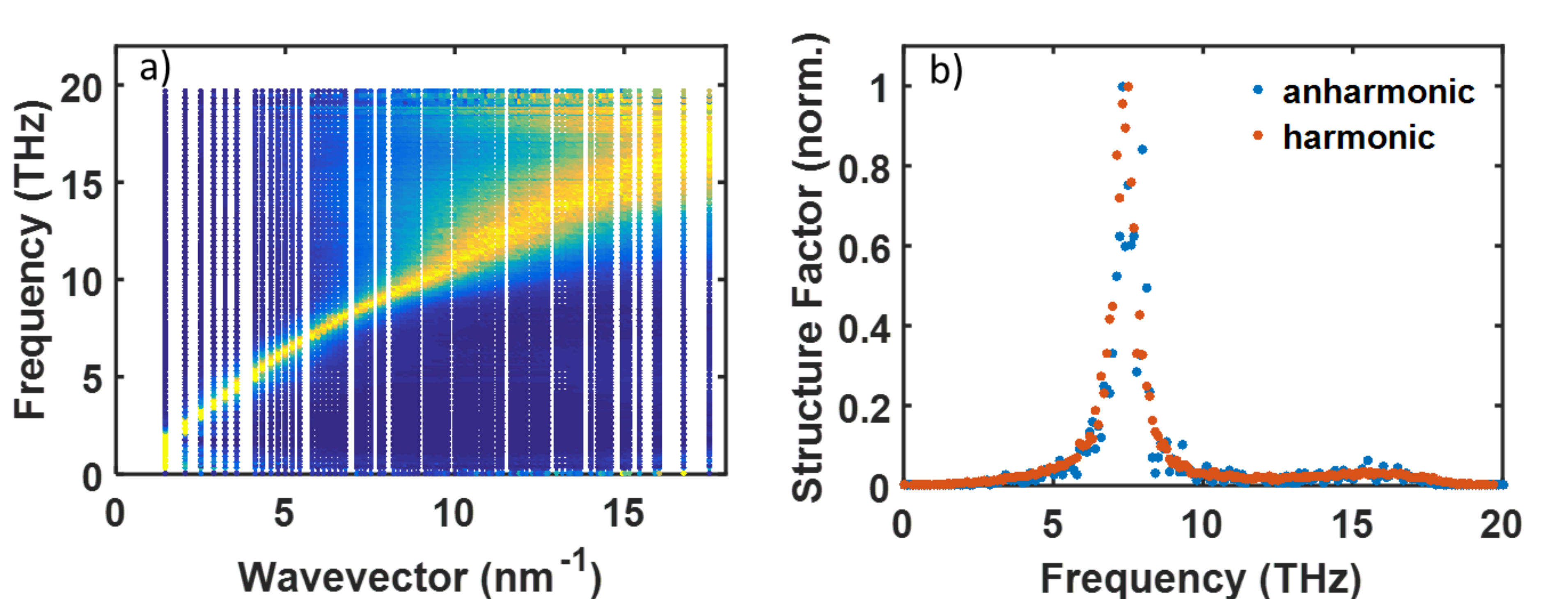}
	\caption{(a) Dynamical structure factor for longitudinal waves for 4096-atom pure a-Si domain. Bright yellow indicates a high intensity of vibrations with the given frequency and wavevector. A clear phonon band is observed up to around 10 THz despite the atomic disorder.  (b)  Constant wavevector slice of the dynamical structure factor at $q$ = 6.0 nm\textsuperscript{-1}. Anharmonic broadening is negligible at 300 K.
	}
	\label{fig:a_Si_part1.pdf}
\end{figure}

	We begin our analysis to gain more insight into the vibrations carrying heat by characterizing the propagating nature of the normal modes of vibration of the amorphous domain. A convenient metric for this characterization is the dynamic structure factor, given by 
	
	\begin{equation}
	S_{L,T}(\boldsymbol{q},\omega) = \sum_{\nu} E_{L,T}(\boldsymbol{q},\nu)\delta(\omega-\omega(\boldsymbol{q = 0},\nu))
	\label{eq:SF_har}
	\end{equation} 	
	where the $\boldsymbol{q}$ is phonon wavevector, $\omega$ is frequency and the summation is over all the modes $\nu$ at $\Gamma$. $E_L$ and $E_T$ refer to the longitudinal polarization and transverse polarization and are defined as 
	\begin{equation}
	E_{L}(\boldsymbol{q}, \nu) = \left|\sum_i [\hat{\boldsymbol{q}}\cdot\boldsymbol{e}(\nu,i)]e^{i\boldsymbol{q}\cdot\boldsymbol{r}_i}\right|^2 
	\end{equation}
	\begin{equation}	
	E_{T}(\boldsymbol{q}, \nu) = \left|\sum_i [\hat{\boldsymbol{q}}\times\boldsymbol{e}(\nu,i)]e^{i\boldsymbol{q}\cdot\boldsymbol{r}_i}\right|^2
	\end{equation}
where the summation is over all atoms indexed by i in the domain, $\hat{\boldsymbol{q}}$ is a unit vector, $\boldsymbol{e}(\nu,i)$ is the eigenvector, and $\boldsymbol{r}_i$ are the equilibrium positions. The dynamic structure factor is precisely what is measured in scattering experiments to measure dispersion relations in crystals and can be applied to search for plane waves in disordered media.

	We calculated the eigenvectors of the 4096 atom structure using the General Utility Lattice Program (GULP) with equilibrated structures from MD \cite{gale_gulp:_1997}.  As amorphous Si is isotropic, we average the dynamic structure factor over all wavevectors of the same magnitude. If propagating waves exist despite the atomic disorder, the dynamic structure factor will exhibit a clear phonon band with a dispersion; if propagating waves are not supported, the vibrational modes will appear diffuse without an apparent dispersion.
	
	The dynamic structure factor for longitudinal waves is presented in Fig. \ref{fig:a_Si_part1.pdf} (a). Unexpectedly, the figure demonstrates that despite the atomic disorder a clear dispersion exists up to frequency as high as 10 THz for longitudinal waves, corresponding to a wavelength of 6.5 \AA. In the transverse direction, a clear dispersion with broadening is also observed up to $\sim 5$ THz, with a similar transition wavelength at 6.6 \AA \space (not shown). For sufficiently high frequency vibrations with wavelengths comparable to interatomic distances, the structure factor is very broad and identifying plane waves with definite frequency and wavevector is not possible. However, the figure clearly shows that propagating elastic waves comprise a substantial portion of the vibrational spectrum. Specifically, by calculating the density of states of the low frequency vibrations with a Debye model, we estimate that about 24\% of all modes are propagating waves. Our observation is consistent with a prior calculation of dynamical structure factor \cite{beltukov_boson_2016} but contrasts with prior studies that restrict propagons to frequencies less than  $2\sim3$ THz in amorphous silicon or comprise a small fraction of all modes  \cite{allen_diffusons_1999,he_heat_2011,larkin_thermal_2014,lv_direct_2016,seyf_method_2016,lv_examining_2016}.

	In addition to the existence of well-defined plane-waves with definite frequency and wavevector, we also observe that the lines are not narrow but have a clear broadening indicating the presence of a scattering mechanism. In crystals, this broadening is typically due to anharmonic interactions. In the perfectly harmonic amorphous solid considered here, anharmonic interactions cannot play any role. Instead, the broadening must be due to fluctuations of the local elastic modulus on length scales comparable to the wavelengths of the propagating waves. Therefore, the picture that emerges from our calculation of dynamical structural factor of a-Si is a vibrational spectrum that is dominated by elastic waves that are scattered by elastic fluctuations in the disordered solid.

	In actual a-Si, anharmonic interactions may increase the scattering rate and hence the broadening. To assess how broadening due to elastic fluctuations compares to that from anharmonic interactions, we also calculate dynamic structure factors using velocity outputs from MD at 300 K \cite{shintani_universal_2008}. The longitudinal dynamic structure factors at $q = 6.0$ nm\textsuperscript{-1} with harmonic and anharmonic forces are depicted in Fig. \ref{fig:a_Si_part1.pdf} (b), demonstrating that the two are nearly identical. Therefore, anharmonic broadening has essentially no effect on the lifetimes and the broadening is solely due to elastic fluctuations.

We next aim to extract quantitative information from the observed broadened lines. Prior works used normal mode analysis to extract lifetimes from molecular dynamics simulations \cite{he_heat_2011,larkin_thermal_2014}. Here, we instead obtain lifetime information for those modes with a well-defined dispersion by fitting a constant wavevector slice of the dynamic structure factor with damped harmonic oscillator (DHO) model \cite{taraskin_low-frequency_1999,shintani_universal_2008,barkema_high-quality_2000,damart_nanocrystalline_2015,beltukov_boson_2016}. The lifetime $\tau$ at a certain frequency is related to the full-width at half-maximum $\Gamma$ by $\tau=1/\pi\Gamma$. By multiplying the lifetimes by the sound velocity, we also obtain mean free paths.

\begin{figure}
	\centering
	\includegraphics[width=1\linewidth]{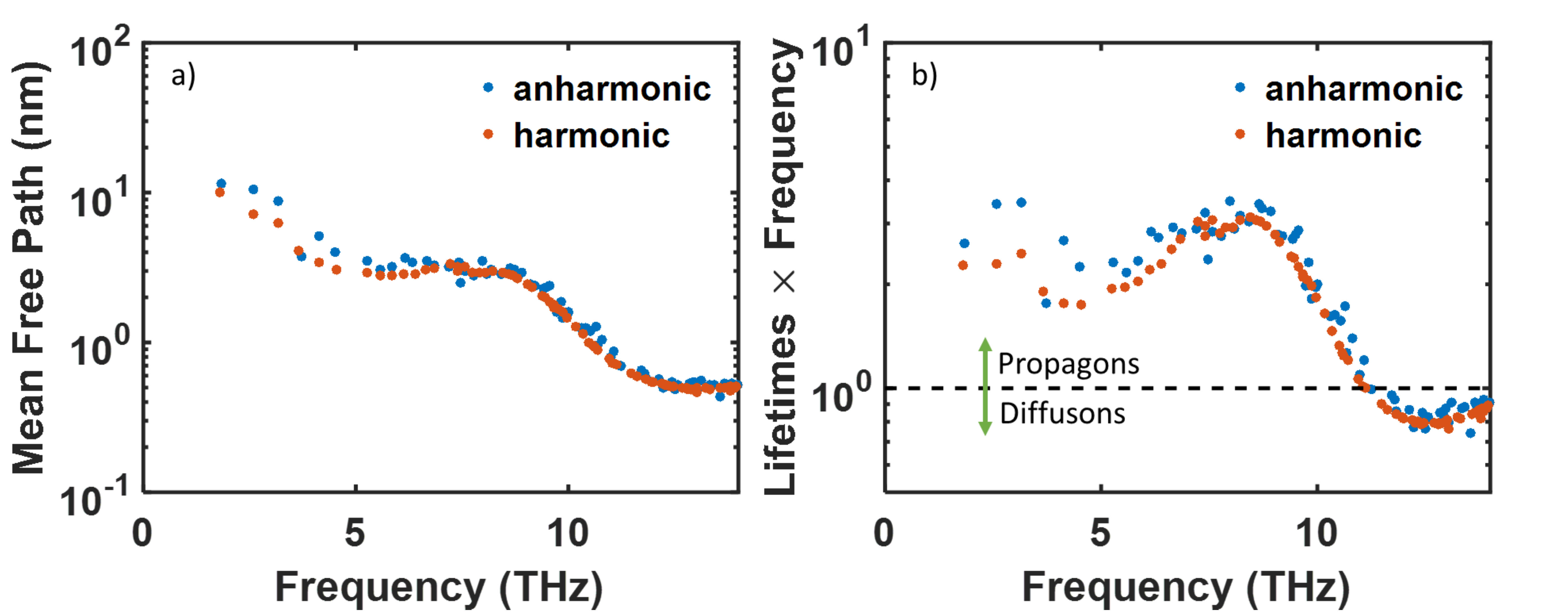}
	\caption{(a) Spectral mean free path and (b) lifetime multiplied by frequency versus frequency for longitudinal waves with harmonic and anharmonic forces for the 4096-atom pure a-Si domain. The Ioffe-Regel criterion occurs when lifetime multiplied by frequency equals 1. Propagons are observed up to around 10 THz for longitudinal waves as predicted from the dispersion. 
	}
	\label{fig:a_Si_part2.pdf}
\end{figure}

	The results are shown in Fig. \ref{fig:a_Si_part2.pdf} (a). We see that the mean free paths span from 0.5 nm to 10 nm. At still lower frequencies that cannot be included in the present simulations mean free paths are likely even longer, as suggested by experiment \cite{kwon_unusually_2017}. In addition, Fig. \ref{fig:a_Si_part2.pdf} (b) plots the product of lifetime and vibrational frequency. In this plot, the Ioffe-Regel (IR) crossover from propagons to diffusons, defined as when the lifetime is equal to the period of a wave, can be indicated as a horizontal line \cite{taraskin_low-frequency_1999}.  For longitudinal waves, the IR crossover is observed at $\sim$ 10 THz. A similar analysis for transverse waves indicates the IR crossover is found around $\sim$ 5 THz for these vibrations; both of these values are in good agreement with the qualitative estimate of the transition frequency from the structure factor. 
	
Having established that propagons comprise a substantial fraction of the vibrational spectrum, we next estimate the propagon contribution to thermal conductivity given knowledge of the linear, isotropic dispersion, the group velocity, and the mean free paths from Figs. \ref{fig:a_Si_part1.pdf} and \ref{fig:a_Si_part2.pdf} using a Debye model. In this model, we separate the propagon contribution into longitudinal and transverse modes with group velocities obtained from the dispersion as 8000 and 3610 m/s, respectively.  Recalling the bulk thermal conductivity of 1.5 Wm\textsuperscript{-1}K\textsuperscript{-1} from the Green-Kubo calculation, our Debye model estimates that propagons contribute about 1.35 Wm\textsuperscript{-1}K\textsuperscript{-1}, or 90 \% of the thermal conductivity. This contribution is much larger than the values reported previously and suggests that, counterintuitively, heat transport in a-Si is dominated by propagating waves despite the atomic disorder. The primary uncertainty in this estimate is the role of vibrations of frequency less than 2 THz that are challenging to include in both the Green-Kubo and structure factor calculations; however, the general conclusion that propagons dominate the heat transport will still hold even in the absence of these additional propagating vibrations.

	\begin{figure*}
		\centering
		\includegraphics[width=1\linewidth]{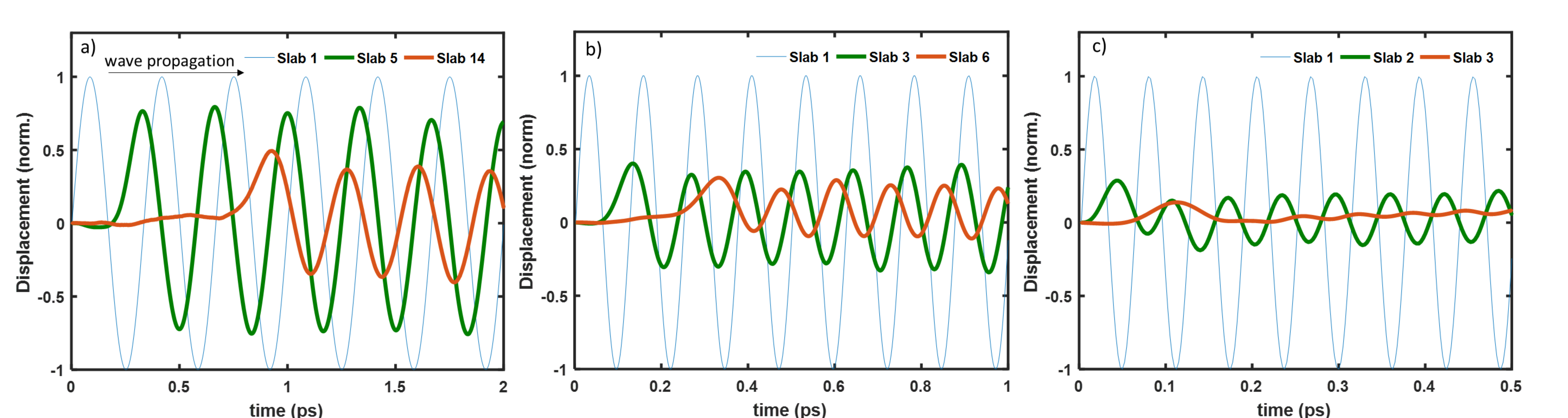}
		\caption{Temporal displacement of atoms in each slab with triggering frequencies a) 3 THz, b) 8 THz, and c) 16 THz. Each sinusoidal wave represents the averaged displacements of the atoms in a slab. By observing where the amplitudes of the displacement decrease by $1/e$, we estimate that the mean free paths are around 9 and 2 nm for 3 THz and 8 THz waves, respectively. The mean free path of the 16 THz wave is comparable to the interatomic spacing and hence the vibration is non-propagating. 
				}
		\label{fig:wavepacket.pdf}
	\end{figure*}
Our analysis thus suggests a considerably different picture of atomic vibrations in a-Si from the conventional view. Earlier works have concluded that propagons are a small fraction of mode population that contribute less than half of the total thermal conductivity, and that the phonon gas model is an inaccurate picture of the vibrations in amorphous solids. By contrast, our examination of the dynamical structure factor indicates that propagons form 24 \% of the mode population and are scattered by fluctuations of the local elastic modulus rather than anharmonicity. Additionally, the highest reported contribution of propagons to thermal conductivity is 50\% of the total. However, the picture that emerges from our analysis is that a gas of delocalized elastic vibrations exists in a-Si, despite the lack of atomic order, which transports most of the heat.

We provide further support for our conclusions with two additional calculations. First, we explicitly demonstrate the propagating nature of certain vibrational modes by conducting a "tuning fork experiment" in which imposed oscillatory atomic motions at one edge of the atomic domain triggers the formation of a traveling wave through the a-Si. To perform this calculation, we first create a domain by repeating 4096-atom cell 10 times along one direction, resulting in a supercell of size 4.3 $\times$ 4.3 $\times$ 43 nm. In the long dimension, the domain is divided into 80 slabs of width 5.431 \AA. Periodic boundary conditions are applied and the temperature is set at 0.1 K to avoid additional thermal displacements. The calculation begins by rigidly displacing the first slab in the longitudinal direction for 2 ps with a sinusoidal wave with amplitude 0.01 \AA \space and a specified frequency. We computed the longitudinal displacements of every atom for time durations less than 2 ps to prevent edge effects, and subsequently averaged the atomic displacements within each slab.

The wave propagation in a-Si at different frequencies is shown in Fig. \ref{fig:wavepacket.pdf}. It is apparent that waves do indeed propagate through a-Si at 3 THz and 8 THz as predicted by the dynamic structure factor calculations. By identifying the location at which the wave amplitude has decreased to $1/e$ of its original value, we estimate that the mean free paths of the 3 THz and 8 THz waves are around 9 nm and 2 nm, respectively. These mean free paths coincide reasonably well with those from dynamic structure factor calculations (8 nm and 3 nm). On the other hand, the excited wave at 16 THz is damped very quickly, and by the second slab the amplitude is already less than $1/e$ of the original value. This observation indicates that at 16 THz the vibration is non-propagating. Therefore, the "tuning fork experiment" qualitatively confirms that propagating waves exist up to a high frequency of around 10 THz in a-Si.

Second, we examine how the thermal conductivity is affected by the partial elimination of elastic fluctuations. Our calculations indicate that fluctuations in local elastic modulus are the dominant mechanism for disrupting propagons. If this assertion is true, we should observe a marked increase in thermal conductivity when these fluctuations are partially eliminated along with a temperature dependence of thermal conductivity that reflects the renewed dominance of phonon-phonon interactions. To test this hypothesis, we generated two additional domains designed to possess reduced elastic fluctuations consisting of 512 and 64-atom amorphous unit cell (AUC) tiled to create 4096-atom structures. The 512 and 64 AUC domains were created using the same melt-quench procedure described earlier. Elastic fluctuations over a length scale equal to the AUC domain size should be eliminated because the same unit cell is tiled repeatedly in space to form the 4096 atom final structure.

\begin{figure}
	\centering
	\includegraphics[width=0.9\linewidth]{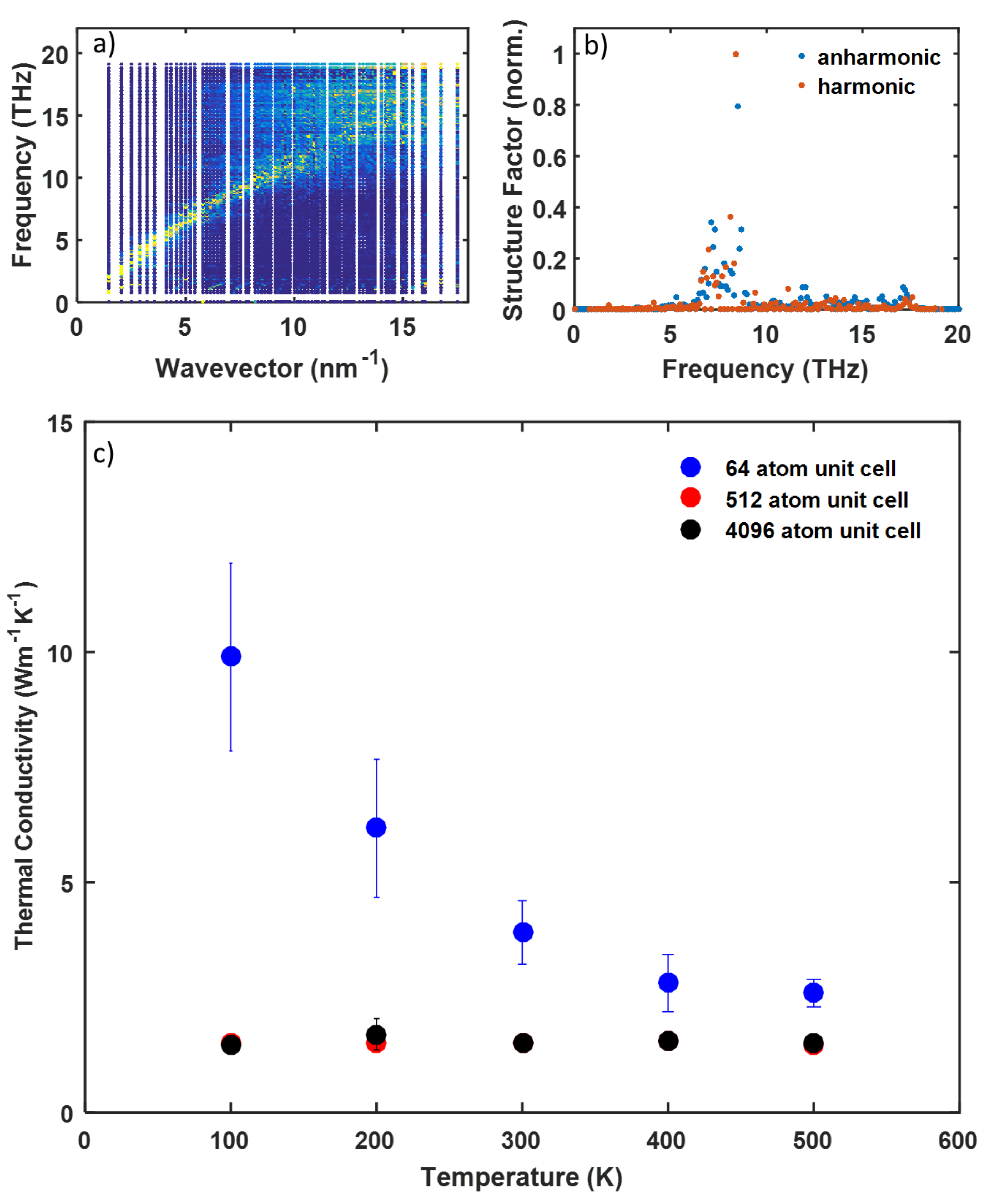}
	\caption{(a) Dynamic structure factor for longitudinal vibrations for the 64 AUC tiled structure. (b) Constant wavevector slice of the dynamical structure factor at $q$ = 6.0 nm\textsuperscript{-1} with harmonic and anharmonic forces for the 64 AUC structure. Anharmonicity plays a role in the broadening. (c) Thermal conductivity versus temperature for three amorphous structures. No temperature dependence is observed for 512 and 4096-atom AUC structures while a noticeable dependence in temperature for the 64-atom AUC tiled structure is evident. 
	}
	\label{fig:Fig4.pdf}
\end{figure}

We followed the same procedure as described earlier to obtain dynamic structure factors for the tiled structures. The structure factor for the 512-atom AUC tiled structure appear almost identical to that of the original calculation (not shown). That for the 64-atom AUC tiled structure are shown in Fig. \ref{fig:Fig4.pdf} (a). We observe discrete points rather than a continuous broadening, indicative of the dynamic structure factor having delta-function-like peaks as occurs in c-Si. From a constant wavevector slice of the dynamic structure factor for the 64-atom AUC tiled structure in Fig. \ref{fig:Fig4.pdf} (b), we observe that anharmonicity broadens the peaks in the frequency range from 5 to 10 THz, indicating that anharmonicity plays a role in scattering these modes. In addition, we observe several peaks in the medium to high frequency region from 10 to 20 THz, which is characteristic of crystalline materials with a multi-atom unit cell that have several folded branches. Overall, these calculations indicate that the 64-atom AUC structure possesses vibrations that are characteristic of a semi-crystalline solid while the 512-atom AUC remains effectively amorphous. Therefore, the key length scale for the elastic fluctuations that sets whether vibrations propagate is around 10 \AA, or the side length of the 64-atom domain.


We now compute the thermal conductivity of the 3 structures using Green-Kubo theory averaged over 10 different initial conditions. The resulting thermal conductivity calculations of these structures are shown in Fig. \ref{fig:Fig4.pdf} (c). The figure shows that the pure a-Si and the 512-atom AUC tiled structure have identical thermal conductivity with little temperature dependence. This result confirms that the 512-atom AUC structure is effectively amorphous. However, we observe a significant increase in thermal conductivity of the 64-atom AUC tiled structure, by more than a factor of 2 at room temperature, along with a marked temperature dependence. At 100 K, the thermal conductivity of the 64-atom AUC tiled structure is $\sim$ 10 Wm\textsuperscript{-1}K\textsuperscript{-1}, more than 6 times that of pure a-Si. Therefore, the 64-atom AUC tiled structure exhibits characteristics of crystals, and the key disorder length scale that sets the transition of thermal vibrations from crystalline to amorphous character lies between 10-20 \AA. 

The picture of a gas of delocalized elastic vibrations transporting heat in amorphous solids suggests follow-on experiments as well as new strategies to realize exceptional thermal materials. First, our prediction of propagons existing up to around 10 THz can be verified with additional thermal measurements on amorphous nanostructures with characteristic dimensions of less than 10 nm as well as with scattering methods such as inelastic X-ray scattering. Second, our analysis suggests that fully dense solids with exceptionally low thermal conductivity can be achieved by manipulating elastic fluctuations, expanding the physical range of thermal conductivity of solids.

In summary, we have examined the atomic vibrations in a-Si using lattice and molecular dynamics calculations. Our study reveals a qualitatively different view of atomic vibrations in a-Si from conventional one in which propagating elastic waves dominate the thermal conduction and are scattered by elastic fluctuations instead of anharmonicity. Our work provides important insights into the long-standing problem of thermal transport in disordered solids.

This work was supported by the Samsung Scholarship, NSF CAREER Award CBET 1254213, and Boeing under Boeing-Caltech Strategic Research and Development Relationship Agreement. 
		
\clearpage	
\bibliography{mylib}	
	
\end{document}